\documentclass[12pt]{article}
\usepackage{a4wide}
\usepackage{xcolor}
\usepackage{amssymb}
\usepackage{amsmath}
\begin{document}
{\renewcommand{\thefootnote}{\fnsymbol{footnote}}
%\hfill  IGC--yy/m--n\\
%\medskip
\begin{center}
{\LARGE  Covariance in models of loop quantum gravity: Spherical symmetry}\\
\vspace{1.5em}
Martin Bojowald,$^1$\footnote{e-mail address: {\tt bojowald@gravity.psu.edu}}
 Suddhasattwa Brahma$^1$\footnote{e-mail address: {\tt sxb1012@psu.edu}} and
Juan D.~Reyes$^2$\footnote{e-mail address: {\tt jdrp75@gmail.com}}
\\
\vspace{0.5em}
$^1$ Institute for Gravitation and the Cosmos,\\
The Pennsylvania State
University,\\
104 Davey Lab, University Park, PA 16802, USA\\
\vspace{0.5em}
$^2$ Centro de Ciencias Matem\'aticas,
Unidad Morelia,\\
 Universidad Nacional Aut\'onoma de
M\'exico, UNAM-Campus Morelia,\\
 A. Postal 61-3, Morelia, Michoac\'an 58090,
Mexico \\
\vspace{1.5em}
\end{center}
}

\setcounter{footnote}{0}

\begin{abstract}
  Spherically symmetric models of loop quantum gravity have been studied
  recently by different methods that aim to deal with structure functions in
  the usual constraint algebra of gravitational systems. As noticed by Gambini
  and Pullin, a linear redefinition of the constraints (with phase-space
  dependent coefficients) can be used to eliminate structure functions, even
  Abelianizing the more-difficult part of the constraint algebra. The
  Abelianized constraints can then easily be quantized or modified by putative
  quantum effects. As pointed out here, however, the method does not
  automatically provide a covariant quantization, defined as an anomaly-free
  quantum theory with a classical limit in which the usual (off-shell) gauge
  structure of hypersurface deformations in space-time appears. The
  holonomy-modified vacuum theory based on Abelianization is covariant in this
  sense, but matter theories with local degrees of freedom are not. Detailed
  demonstrations of these statements show complete agreement with results of
  canonical effective methods applied earlier to the same systems (including
  signature change).
\end{abstract}

\section{Introduction}

Several suggestions have been made in models of loop quantum gravity which may
indicate a potential to provide interesting physical effects. Popular examples
are mechanisms to avoid some of the singularities encountered in classical
general relativity. Following from a crucial step in the procedure of loop
quantization, most of these effects are based on a replacement of polynomial
(extrinsic) curvature expressions in the canonical Hamiltonian of the
classical theory by bounded (and usually periodic) functions. As can easily
be seen by the example of isotropic models, in which the classical
Hubble-squared term in the Friedmann equation would be turned into a bounded
function, it is then not surprising that upper bounds on curvature or energy
densities can be obtained. A more crucial consistency question, also posed in
\cite{SmallLorentzViol}, is whether the resulting modified theories can be
covariant, or whether the upper bounds on curvature amount to a
symmetry-breaking cut-off.

In canonical formulations such as loop quantum gravity, covariance is not
manifest but still plays an important role. Instead of using coordinate
transformations of space-time tensors, canonical theories refer to gauge
transformations which, in geometrical terms, generate deformations of spatial
hypersurfaces in space-time \cite{Regained}. The generator of a deformation
normal to a hypersurface is the above-mentioned gravitational Hamiltonian. If
it is modified by bounded curvature expressions (or other quantum
corrections), it is unclear whether it can still generate gauge
transformations. Mathematically, the question is whether modified Hamiltonians
can retain closed Poisson brackets or commutators with themselves and with
generators of spatial deformations tangential to hypersurfaces. Some
information about this question has been gained in recent years using
effective
\cite{ConstraintAlgebra,ScalarHolInv,JR,LTBII,ModCollapse,HigherSpatial} and
operator methods
\cite{ThreeDeform,TwoPlusOneDef,TwoPlusOneDef2,AnoFreeWeak,SphSymmOp}. Here we
will follow a new but, as we will see, not independent direction toward the
same question.

Covariance cannot be addressed in minisuperspace models such as isotropic
cosmological ones, because they do not show how temporal and spatial
variations of fields are related. The simplest inhomogeneous models are
obtained by imposing spherical symmetry, to be considered in this paper. In
this setting one has a non-trivial set of hypersurface-deformation generators
and brackets or commutators between them. As in the full theory, the bracket
of two normal deformations has structure functions instead of structure
constants, so that the generators do not form a Lie algebra. The usual
quantization methods of gauge theories therefore complicate considerably, and
existing quantizations of spherically symmetric models use either
reformulations of the generators \cite{SphKl1} or quantize the reduced phase
space from which the gauge flow has been eliminated \cite{SphKl2,Kuchar}. An
interesting new proposal of reformulating the generators (and at the same time
including some ingredients of a loop quantization) is the Abelianization of
normal deformations found recently in
\cite{LoopSchwarz,LoopSchwarz2}. Compared with earlier Abelianizations
\cite{Strobl}, an important feature mentioned in \cite{LoopSchwarz2} is that
it works even when a scalar field with local physical degrees of freedom is
included. There is therefore a chance that the problem of structure functions
may be overcome at least in these models.

A question left open in \cite{LoopSchwarz,LoopSchwarz2} is whether the
resulting quantizations are covariant. By quantizing a system in which the
brackets of gauge generators have been turned into a Lie algebra, the
constructions of \cite{LoopSchwarz,LoopSchwarz2} certainly provide consistent
quantum models. However, it is not clear whether or in what sense they are
models of quantum gravity with a consistent space-time picture. This is the
question we turn to in the present paper, starting with a discussion of what
it means for a canonical theory to be covariant. We will show that the
loop-modified vacuum model of \cite{LoopSchwarz} is covariant only if the
original Hamiltonian, prior to Abelianization, is modified in a restricted way
with exactly the same conditions found by effective methods \cite{JR}. There
is therefore a remarkable convergence between results of Abelianization and
the effective framework. We will also show that the modified model of
\cite{LoopSchwarz2} with a scalar field is not covariant, unless a background
treatment is used for the scalar on a vaccum solution so that matter and
gravity have non-matching versions of covariance. More broadly, we point out
that to date no covariant inhomogeneous model with local physical degrees of
freedom has been found with holonomy modifications from loop quantum gravity
(while such models exist for curvature-independent inverse-triad corrections).

\section{Covariance in canonical terms}

The canonical formulation of general relativity leads to a phase space given
by the spatial metric $q_{ab}$ and momenta related to extrinsic curvature
$K_{ab}$. It is subject to the Hamiltonian constraints $H[N]$, labelled by
spatial lapse functions $N$, and diffeomorphism constraints $D[M^a]$, labelled
by spatial shift vector fields $M^a$. These constraints are first class with
closed brackets \cite{DiracHamGR,ADM}
\begin{eqnarray}
 \{D[M_1^a],D[M_2^a]\} &=& D[{\cal L}_{M_1}M_2^a] \label{DD}\\
 \{H[N],D[M^a]\} &=& -H[{\cal L}_MN] \label{HD}\\
 \{H[N_1],H[N_2]\} &=& D[q^{ab}(N_1\partial_bN_2-N_2\partial_bN_1)]\,. \label{HH}
\end{eqnarray}
They generate gauge transformations representing hypersurface deformations
\cite{Regained}. On the space of solutions to the constraints, the same gauge
transformations are equivalent to Lie derivatives along space-time vector
fields, and therefore represent coordinate freedom. Manifest covariance is
replaced by gauge covariance under hypersurface deformations. (For more
details on canonical gravity, see for instance \cite{CUP}.)

\subsection{Conditions}

This well-known result leads us to two conditions to be realized for a
modified or quantized canonical theory to be covariant:
\begin{enumerate}
\item[(i)] The classical generators $H[N]$ and $D[M^a]$ must be replaced by
  generators which still have closed brackets, computed either as Poisson
  brackets in a modified or effective theory, or as commutators of operators
  in a quantization.
\item[(ii)] Brackets of the new generators of gauge transformations must have
  a classical limit identical with the classical brackets
  (\ref{DD})--(\ref{HH}).
\end{enumerate}
When condition (i) is satisfied, one has a consistent gauge theory since the
gauge generators eliminate the same number of spurious degrees of freedom as
in the classical case. But only when conditions (i) and (ii) are satisfied
does one have a consistent {\em space-time theory}, in which there is a
classical regime with the correct space-time structure. Accordingly, we call a
modified, effective, or quantum theory {\em covariant} if and only if
conditions (i) and (ii) are satisfied. The constructions in
\cite{LoopSchwarz,LoopSchwarz2} have provided consistent gauge theories
obeying (i), but the question of covariance or condition (ii) has not been
addressed yet.

An important aspect of condition (ii) is that it is an off-shell statement,
for which not only the solution space of constraints $H[N]=0$ and $D[M^a]=0$
is relevant but also the behavior of fields not satisfying the
constraints. This dependence on off-shell properties is in agreement with
the usual understanding of space-time covariance, in which one makes use of
line elements or metric tensors not necessarily solving Einstein's (or
modified) field equations. It is also an important part of our classical
picture of space-time as a stage on which different matter systems can be set
up. Even though space-time and matter interact with each other, the covariance
conditions commonly posed for matter theories require certain symmetries of
the action on {\em any} background space-time, not necessarily one solving
Einstein's equation. The usual covariance statements about (classical or
quantum) matter systems on a classical space-time are therefore off-shell. For
all we know, there could well be stronger interrelations between space-time
and matter if both ingredients are quantum, so that it would no longer be
possible to separate a covariant matter theory from an anomaly-free
space-time. However, for the combined system to have the correct classical
limit, our condition (ii), which is formulated only in this limit, must still
hold.

\subsection{Background treatment}

In this context, one should therefore avoid taking the viewpoint that on-shell
properties are sufficient to decide whether a space-time theory is
meaningful. Although all observables computed with a given solution refer to
the constraint surface modulo gauge transformations, covariance in the form
usually used is a statement about a partial solution space. (For additional
reasons, see \cite{NPZRev}.)  Moreover, the full solution space of general
relativity or a modified version is too unwieldy and in many cases of interest
does not allow manageable on-shell statements in complete terms. Even
models such as spherically symmetric gravity with a scalar field remain
challenging in this setting. Most evaluations of gravitational theories
(including \cite{LoopSchwarz2}) make use of some kind of background
approximation, in which one starts with a simple-enough vacuum solution and
then perturbatively includes additional inhomogeneity or matter fields on this
background. In practice, the background picture is therefore even more
pronounced than the conceptual discussions of the preceding paragraph might
indicate.

In more technical details, consistency of a matter model as a space-time
theory may be formulated by requiring the fields to satisfy the local
conservation equation $\nabla^{\mu}T_{\mu\nu}=0$ for their stress-energy
tensors. Canonically, as shown in \cite{Energy}, this equation follows from
the analogs of (\ref{DD})--(\ref{HH}) for a matter Hamiltonian (assuming, for
simplicity, that no curvature couplings are present).  In particular, relating
stress-energy components to different kinds of derivatives of the matter
contributions ${\cal H}_{\rm matter}$ and ${\cal D}_a^{\rm matter}$ to the
local constraints, one can derive the equation
\begin{eqnarray}
N\sqrt{\det q}\:\nabla_{\mu}T^{\mu}{}_0 &=& -N\frac{\partial {\cal H}_{\rm
    matter}}{\partial
   t}- N^a \frac{\partial {\cal
      D}_a^{\rm matter}}{\partial t}\\
&& + {\cal L}_{\vec{N}} C_{\rm matter}[N,N^a] +
  \frac{\partial q_{ab}}{\partial t} \frac{\delta H_{\rm matter}}{\delta
    q_{ab}}\nonumber \\
 && +\partial_b\left(N^2q^{ab}{\cal D}_a^{\rm matter}+
 2N^cq^{ba}
   \frac{\delta H_{\rm matter}}{\delta q^{ac}}\right)\,. \nonumber
\end{eqnarray}
(The total matter contribution, summing the smeared contributions to the
Hamiltonian and diffeomorphism constraints, is denoted by $C_{\rm
  matter}[N,N^a]$.)  The classical off-shell brackets (and not just closed
constraint brackets of some form) imply that the two terms $\partial{\cal
  H}_{\rm matter}/\partial t=\{{\cal H}_{\rm matter},H[N,N^a]\}$ and
$\partial^a(N^2{\cal D}_a^{\rm matter})$ cancel out if (\ref{HH}) holds, and
the rest is zero based on other identities.  A conservation law therefore
follows only if the brackets are not just closed but (in the classical limit)
of precisely the form obtained for the classical hypersurface
deformations. (In \cite{Energy}, a matter Hamiltonian without curvature
coupling has been assumed for simplicity, in which case the matter Hamiltonian
and diffeomorphism generators alone have brackets of the form
(\ref{DD})--(\ref{HH}). Again, the importance of off-shell properties is
underlined because the matter contributions to the constraints need not vanish
separately. Some quantum effects, like those to be studied in the rest of this
paper, may introduce additional curvature couplings, but they disappear in the
classical limit in which the off-shell condition (ii) is formulated.)

For these independent reasons, off-shell brackets are relevant in the
definition of covariance and should be checked before one can claim that a
quantized model is a quantum theory of space-time. Even if one uses a
background treatment for a matter field on a vacuum solution which latter has
been shown to be covariant, there are still conditions to be imposed on the
matter model: the existence of a local conservation law. A background
treatment makes the construction of models less restrictive, but still such a
procedure is far from being arbitrary.

The difference between a background treatment and a background-independent
model in standard formulations is that only the latter ensure the existence of
solutions to the coupled equations of gravity and matter, such as
$G_{\mu\nu}=8\pi G T_{\mu\nu}$ for general relativity. Compared to a
background treatment, coupling gravity to matter in a consistent way implies
additional restrictions even if the coupled equations are not actually solved,
that is if no back-reaction is considered.  Classically, the equation is
consistent because the contracted Bianchi identity for $G_{\mu\nu}$ and the
local conservation law for $T_{\mu\nu}$ take the same form.

In models of loop quantum gravity, the contracted Bianchi identity, in its
canonical form as Poisson brackets of gravitational constraints, is
generically modified. Instead of (\ref{HH}), we usually have
\begin{equation} \label{HHbeta}
 \{H[N_1],H[N_2]\} = D[\beta q^{ab}(N_1\partial_bN_2-N_2\partial_bN_1)]
\end{equation}
with a phase-space function $\beta$ depending on the spatial metric $q_{ab}$
or extrinsic curvature.  A consistent background-independent model then
requires the local conservation law, or the Poisson brackets of matter
contributions to the constraints, to be modified in a matching way with the
same function $\beta$. (We emphasize again that this condition is important
even if back-reaction of matter on space-time is not considered by solving the
coupled equations.) A background treatment, on the other hand, merely requires
that the gravitational brackets and matter brackets have consistent but not
necessarily matching forms. These contributions would both obey
(\ref{HHbeta}), but possibly with different functions $\beta$ for gravity and
matter. As background models, such theories would still be formally
consistent, but it would not be clear whether they could be background
formulations of covariant background-independent models. The quantization
proposed in \cite{LoopSchwarz2} is an example for a background treatment
which, as demonstrated by the derivations that follow, is not known how to be
embedded in a covariant background-independent theory of the same symmetry
type (setting aside the vastly more complicated question of embedding it in
some full quantum theory).

\section{Abelianization of normal deformations in spherically symmetric
  models}

Compared with \cite{SphKl1,Kuchar}, the formulation of spherically symmetric
models with real connection variables, given in \cite{SphSymm} is most
relevant for the inclusion of loop effects as they are currently
understood. We first recall these variables for notational purposes, and then
discuss features of constraints and possible modifications.

\subsection{Classical theory}

Using a radial variable $x$, not
necessarily identical to the area radius $r$, the spatial metric or line
element
\begin{equation}
 {\rm d}s^2 = \frac{(E^{\varphi})^2}{|E^x|}{\rm d}x^2+ |E^x|({\rm
   d}\vartheta^2+\sin^2\vartheta{\rm d}\varphi^2)
\end{equation}
is expressed by two functions $E^x(x)$ and $E^{\varphi}(x)$ which are the
independent components of a densitized triad reduced to spherical symmetry
\cite{SymmRed}. (While $E^x$ is a 1-dimensional scalar in the reduced model,
$E^{\varphi}$ has density weight one; see \cite{SphSymm}.) The triad
components are canonically conjugate to components of extrinsic curvature:
\begin{equation}
 \{K_x(x),E^x(y)\}=G\delta(x,y) \quad,\quad \{K_{\varphi}(x),E^{\varphi}(y)\}=
 \frac{1}{2}G\delta(x,y)\,.
\end{equation}

\subsubsection{Vacuum model}

The reduced diffeomorphism constraint has only one component,
\begin{equation}\label{Diffeoclassical}
 D[M] = \frac{1}{G} \int{\rm d}x M(x) \left(-\frac{1}{2}(E^x)'K_x+K_{\varphi}'
   E^{\varphi}\right)\,,
\end{equation}
and the reduced Hamiltonian constraint is
\begin{equation}
 H[N]=-\frac{1}{2G}\int{\rm d}x N(x) \left(|E^x|^{-\frac{1}{2}}
   E^{\varphi}K_{\varphi}^2+
2 |E^x|^{\frac{1}{2}} K_{\varphi}K_x
+ |E^x|^{-\frac{1}{2}}(1-\Gamma_{\varphi}^2)E^{\varphi}+
2\Gamma_{\varphi}'|E^x|^\frac{1}{2}\right)
\end{equation}
with the spin-connection component $\Gamma_{\varphi}=-(E^x)'/2E^{\varphi}$.
It is a lengthy but straightforward exercise to confirm that these phase-space
functions have the brackets (\ref{DD})--(\ref{HH}) with the inverse spatial
metric $q^{ab}$ replaced by the one component $|E^x|/(E^{\varphi})^2$. These
brackets control covariance in the reduced model, that is covariance under
transformations preserving spherical symmetry.

The reduced model still has structure functions. However, as noted in
\cite{LoopSchwarz}, the linear combination
\begin{equation}
 \tilde{{\cal C}}:= \frac{(E^x)'}{E^{\varphi}} {\cal H}-
 2\frac{K_{\varphi}\sqrt{|E^x|}}{E^{\varphi}} {\cal D}
\end{equation}
of the original local constraints ${\cal H}$ and ${\cal D}$ allows one to
eliminate $K_x$ from the new constraint $\tilde{{\cal C}}$ replacing ${\cal
  H}$ (leaving ${\cal D}$ unchanged). Moreover, in the vacuum case,
$\tilde{\cal C}={\cal C}'$ is a total derivative, so that integration by parts
removes one derivative at the (small) expense of working with a densitized
lapse function $N'=:L$. Since the final smeared constraint
\begin{equation} \label{CL}
 C[L]=\int {\rm d}x L(x){\cal  C}(x) = -\frac{1}{G}
\int{\rm d}x L(x) \left(\sqrt{|E^x|}
 \left(1+K_{\varphi}^2- \Gamma_{\varphi}^2\right)+{\rm const.}\right)\,,
\end{equation}
obtained after integrating by parts $N\tilde{\cal C}=N{\cal C}'$, depends
neither on $K_x$ nor on spatial derivatives of $K_{\varphi}$ or $E^{\varphi}$,
the antisymmetric Poisson bracket of the final constraints $C$ is trivially
zero, while
\begin{equation} \label{CD}
 \{C[L],D[M]\}=C[(ML)']
\end{equation}
as suitable for a constraint with densitized lapse function $L=N'$.

Our Equation (\ref{CD}) corrects a small mistake in Equation (15b) of
\cite{LoopSchwarz3} which has important conceptual ramifications. In
(\ref{CL}), an undetermined constant appears because $C[L]$ is derived only
for $L=N'$ and boundary terms are ignored in \cite{LoopSchwarz}. (The constant
can be related to the classical ADM mass if asymptotic flatness is assumed.)
The presence of a constant, which does not contribute to the left-hand side of
(\ref{CD}), is consistent with (\ref{CD}) because the smearing function
$(ML)'$ on the right-hand side is again a total derivative. This smearing
function (rather than $ML'$) not only follows from a direct calculation of the
bracket, it is also the correct Lie derivative ${\cal L}_{M{\rm d}/{\rm
    d}x}L=ML'+M'L$ of a scalar $L$ of density weight one. (Recall that $L$ is
defined as $N'$, the derivative producing a density weight in the
1-dimensional radial manifold.)

\subsubsection{Scalar field}

With all these features, the original Abelianization of the vacuum constraint
might look special and coincidental. However, as noted rather in passing in
\cite{LoopSchwarz2}, the same basic idea can be used to Abelianize the bracket
of two normal deformations for models with a scalar field, except that the
constraint is no longer a total derivative and one does not integrate by
parts: The analog of the previous smeared $\tilde{\cal C}$ is now
\begin{eqnarray}\label{MattHam}
 C[N]&=& \frac{1}{G}
 \int{\rm d}x N(x) \Biggl(-\frac{1}{2} \frac{(E^x)'}{\sqrt{|E^x|}}
     (1+K_{\varphi}^2)- 2\sqrt{|E^x|} K_{\varphi}K_{\varphi}'\\
 &&+
     \frac{(E^x)'}{8\sqrt{|E^x|}(E^{\varphi})^2} \left(4E^x(E^x)''+
       ((E^x)')^2\right) - \frac{1}{2} \frac{((E^x)')^2 \sqrt{|E^x|}
       (E^{\varphi})'}{(E^{\varphi})^3}\nonumber\\
&& + 2\pi G
     \frac{(E^x)'}{\sqrt{|E^x|}(E^{\varphi})^2} \left(P_{\phi}^2+(E^x)^2
       (\phi')^2\right)- 8\pi G\sqrt{|E^x|} \frac{K_{\varphi}}{E^{\varphi}}
     P_{\phi}\phi'\Biggr)\,. \nonumber
\end{eqnarray}
The Abelianization property is not trivial at all, but by an explicit
calculation one can confirm that it is still true.  As we shall see, it
generalizes to other matter fields as well.

There is therefore a chance that Abelianizations of normal deformations can
give rise to generic results at least in midi-superspace models. (Indeed,
normal deformations in polarized Gowdy models can be Abelianized in a very
similar way \cite{GowdyCov,GowdyAbel}.) Since the method relies on eliminating
one component of extrinsic curvature from the Hamiltonian constraint, it is
not clear how useful it could be in the full theory where no component is
distinguished. It is also important that $H$, like $D$, is linear in the
extrinsic-curvature component to be eliminated, which again is not true for
any component in the full theory.

\subsection{Modifications}

Loop quantization of spherically symmetric models \cite{SphSymm,SphSymmHam}
proceeds by turning $E^x$ and $E^{\varphi}$ into derivative operators on
spin-network states, while $K_x$ and $K_{\varphi}$ are not directly
represented. Instead, these degrees of freedom are realized via holonomy
operators quantizing $h_{[x_1,x_2]}:=\exp(i\int_{x_1}^{x_2}K_x(x){\rm d}x)$
and $h_{\{x\}}:= \exp(iK_{\varphi}(x))$. (We label ``extended holonomies'' of
$K_x$ by intervals $[x_1,x_2]$ and ``point holonomies'' of $K_{\varphi}$ by
points $\{x\}$.)  The first expression is a gauge-invariant version of the
U(1)-holonomy of the $x$-component of a connection, while the second
expression models the same exponential behavior for the angular component.

In order to proceed to a quantization of the constraints, one has to make sure
that all ingredients can be expressed by holonomies instead of curvature (or
connection) components. Since the classical constraints are at most quadratic
in the latter, they require modifications (often viewed as regularizations)
before they can be turned into operators. (One can avoid modifications of the
diffeomorphism constraint by representing the finite flow it generates instead
of the infinitesimal generator \cite{ALMMT}. We comment on this step and
possible problems in App.~\ref{a:Diff}.) As mentioned in the introduction,
unbounded functions of the classical curvature components are then replaced by
bounded functions such as $h_{\{x\}}$ for $K_{\varphi}^2$. Applied to the
Hamiltonian constraint, this process amounts to a modification which may break
covariance.

In \cite{LoopSchwarz,LoopSchwarz2}, consistent gauge theories have been found
even with a modification of the $K_{\varphi}$-dependence, making use of
Abelianization results. However, the covariance question remains to be
addressed. We now answer this question (with two different outcomes) in the
two cases of the vacuum model and the scalar model. After this, we extend
Abelianization results to general spherically symmetric matter systems, with
the same outcome as for a scalar field.

\subsubsection{Vacuum model}

It is clear that a modified constraint $C[L]$ obtained after replacing
$K_{\varphi}^2$ in (\ref{CL}) by $\delta^{-2}\sin^2(\delta K_{\varphi})$ (or
any other function of $K_{\varphi}$) preserves the Abelian nature of the
vacuum constraint. Condition (i) for a consistent gauge theory is therefore
respected by the modification. The question whether condition (ii) for a
space-time model is respected is less trivial to answer. Without the
modification, we know that the Abelian constraint comes from a system which
obeys the classical hypersurface-deformation brackets. However, this
observation does not guarantee that there is a formulation of the modified
constrained system which (i) is closed for all values of $\delta$ and (ii) has
brackets in agreement with classical generators of hypersurface deformations
in the classical limit $\delta\to0$.

Let us begin by modifying the first two terms of the usual classical
Hamiltonian constraint with arbitrary functions of the extrinsic-curvature
component $K_\varphi$. This procedure is equivalent to including only
point-wise holonomy corrections for the angular component of the connection
coefficient:
\begin{eqnarray}\label{modHam}
 H[N]&=&-\frac{1}{2G}\int{\rm d}x N(x) \left(|E^x|^{-\frac{1}{2}}
   E^{\varphi}f_1\left(K_{\varphi}\right)+
2 |E^x|^{\frac{1}{2}} f_2\left(K_{\varphi}\right)K_x\right.\\
&&+ \left.|E^x|^{-\frac{1}{2}}(1-\Gamma_{\varphi}^2)E^{\varphi}+
2\Gamma_{\varphi}'|E^x|^\frac{1}{2}\right)\,. \nonumber
\end{eqnarray}
We first define a new linear combination of the modified Hamiltonian
constraint and the usual diffeomorphism constraint, just as in the classical
case, to eliminate $K_x$ from the new constraint while leaving the
diffeomorphism constraint unchanged
\begin{equation}\label{newcnstrnt}
 \tilde{{\cal C}}:= \frac{(E^x)'}{E^{\varphi}} {\cal H}-
 2\frac{f_2\left(K_{\varphi}\right)\sqrt{|E^x|}}{E^{\varphi}} {\cal D}.
\end{equation}
The new constraint has the form
\begin{eqnarray}
 \tilde{C}[N]&=&-\frac{1}{G}\int {\rm d}x N(x)\tilde{{\cal  C}}(x)  \label{tildeC1} \\
  &=& -\frac{1}{G}\int{\rm d}x N(x)\left\{\frac{{\rm d}}{{\rm d}x}\left[
      \sqrt{|E^x|}
 \left(1- \Gamma_{\varphi}^2\right)\right] +
\frac{1}{2}|E^x|^{-1/2}(E^x)'f_1+2|E^x|^{1/2}f_2
K_\varphi^{\prime}\right\}\,. \nonumber
\end{eqnarray}
It is straightforward to see that the condition for $\tilde{{\cal C}}$ to be a
total derivative is
\begin{equation} \label{f2f1}
 2f_2=\frac{{\rm  d}f_1}{{\rm d}K_{\varphi}}\,.
\end{equation}
If this equation is true, we obtain a Lie algebra for the system of
constraints as in the classical case.  A more-general analysis of consistent
modifications of the Abelianized constraints is given in Sec.~\ref{s:Gen}.

Alternatively, we could have started from the classical version of the new
constraint $C[N]$ in (\ref{CL}), after Abelianization, introduced the
modification function $f_1$ as in \cite{LoopSchwarz}, and then asked whether
the modified constraint can be redefined as part of a constrained system with
hypersurface deformations as the classical limit. To do so, we should find out
how we can go from the (modified) Abelianized system of constraints to a new
system of constraints ${\cal H}$ and ${\cal D}$ which, in the classical limit,
are the generators of hypersurface deformations.  After modifying the
Abelianized constraint, we go back to a system of Hamiltonian and
diffeomorphism constraints by a linear combination of ${\cal D}$ with the new
constraint, which can only be the inverse of (\ref{newcnstrnt}), with $f_2$
obeying (\ref{f2f1}) for the correct hypersurface-deformation brackets to be
realized in the classical limit (after integrating by parts the modified
(\ref{CL})).  The new system has the classical diffeomorphism constraint and a
modified Hamiltonian constraint with the first two terms proportional to
functions of $K_\varphi$ which automatically obey the relation (\ref{f2f1}) as
a consequence of integrating by parts.

\subsubsection{Equivalence with effective methods and deformed constraint
  brackets}

This outcome, including the precise form of the relation (\ref{f2f1}), is just
what happens when one tries to close the algebra of constraints without
Abelianization, starting with holonomy modifications directly in the
Hamiltonian constraint \cite{JR,HigherSpatial}. Thus, the Abelianized system
of constraints in \cite{LoopSchwarz,LoopSchwarz2} is equivalent to the system
of modified constraints with deformed structure functions from effective
models, provided one makes sure that the modified system is still
covariant. In particular, the hypersurface-deformation brackets are closed but
deformed for $\delta\not=0$.

Although Abelianization of normal deformations allows one to remove structure
functions from the brackets of constraints, for covariant versions the same
modifications of brackets of hypersurface deformations are obtained as found
in direct treatments of structure functions \cite{JR,HigherSpatial}: For
holonomy-modified spherically symmetric models, we have brackets
(\ref{HHbeta}) with
\begin{equation}    \label{betaK}
 \beta=\frac{\partial f_2}{\partial
   K_{\varphi}}=\frac{1}{2}\frac{\partial^2f_1}{\partial K_{\varphi}^2}
\end{equation}
using (\ref{f2f1}). This function is negative near a local maximum of $f_1$,
indicating signature change \cite{Action,SigImpl}.  This important consequence
and related implications of holonomy modifications cannot be avoided by
reformulating the constraint algebra because covariance conditions still
require one to check the brackets of hypersurface deformations even if their
generators are not used directly as constraints. Realizing these
relationships, there is complete agreement between the modified models based
on Abelianizations, presented in \cite{LoopSchwarz,LoopSchwarz2}, and the
earlier constructions of anomaly-free effective models in
\cite{JR,HigherSpatial}.

\subsubsection{Scalar field}

It is easy to see that the Hamiltonian constraint of a spherically symmetric
gravity theory coupled to matter {\em cannot} be modified according to
holonomy corrections as incorporated previously. If we look back at the
classical form of the Hamiltonian (\ref{MattHam}), we realise that Abelization
works due to some subtle cancellations. The bracket between the second term
from the gravitational part in (\ref{MattHam}) (proportional to
$K_\varphi^{\prime}$) and the first term from the scalar part (proportional to
$P_\phi$) is cancelled by the bracket between the first term and the third
term (proportional to $P_\phi \phi^\prime$), both from the scalar
part. Similarly, the bracket between the same (second) term from the
gravitational part and the second term from the scalar part (proportional to
$\phi^\prime$) is cancelled by the term arising from the bracket between the
second and third term of the scalar part. However, the most interesting
cancellation happens between the brackets of the first and second terms of the
scalar part and the bracket of the fourth term of the gravitational part
(proportional to $(E^{\varphi})^\prime$) and the third term of the scalar
part.

If we now replace the extrinsic-curvature components by some arbitrary
functions of this variable, the resulting bracket of constraints can never be
made to close into a combination of constraints, let alone made zero for an
Abelian bracket. If we replace $K_\varphi^2$ in the gravitational part by some
function $f(K_\varphi)$, then the $K_\varphi$ in the third term of the scalar
part has to be replaced by ${\rm d}f/{\rm d}K_\varphi$, such that the first
two pairs of cancellations are still valid just as in the classical
case. However, with this modification, the bracket between the first and
second terms of the scalar part (which do not contain $K_{\varphi}$ to be
modified) is \textit{not} cancelled by the bracket coming from the term
proportional to $(E^{\varphi})^\prime$ from the gravitational part and the
third term from the scalar part, the latter now having been
modified. (Section~\ref{s:Gen} contains a more-explicit demonstration.)

Although the result is negative in the sense that a simple Abelianization does
not lead to a covariant modified theory, there is again agreement with
effective methods.  Attempts to include scalar fields in spherically symmetric
models within an effective approach, along the lines of
\cite{JR,HigherSpatial} for vacuum models, have failed to provide closed
brackets of constraints including holonomy modifications. The reason for this
lack of closure is the appearance of precisely the same terms that do not
cancel out in an attempted Abelianization. At present, it is not known whether
holonomy-modified spherically symmetric models with a scalar field can be
anomaly-free, or whether their normal deformations can be Abelianized. We will
demonstrate the equivalence of these negative results based on effective
methods and partial Abelianizations after introducing more-general matter
systems.

\subsubsection{General matter model}
\label{s:Gen}

We now consider generic (spherically symmetric) matter systems with
non-derivative couplings to gravity. We assume a consistent or first-class
gravity-matter system of this kind, which has been obtained by inserting
modification functions in a classical matter system without curvature coupling
and higher spatial derivatives. The classical matter Hamiltonian therefore
obeys the bracket (\ref{HH}) on its own, without including gravitational
terms. We assume same property to be true for a modified Hamiltonian obtained
in this way even if modification functions are allowed to depend on curvature
components (but not on spatial derivatives). In fact, it turns out to be
difficult to find consistent modified theories violating this assumption
because cross-terms of the gravitational and matter parts of constraints in
the $\{H,H\}$-bracket would lead to higher spatial derivatives in the bracket
which, if non-zero, could not be absorbed in a constraint to produce a
first-class system One can also confirm this property explicitly for the
matter Hamiltonians given below, where correction functions are allowed to
depend on $K_{\varphi}$.

The form of modifications assumed here therefore implies that the matter parts
of the diffeomorphism and Hamiltonian constraints, $H_{\rm matter}[N]=\int{\rm
  d} x N\mathcal{H}_{\rm matter}$ and $D_{\rm matter}[M]=\int{\rm d} x
M\mathcal{D}_{\rm matter}$, satisfy
\begin{align}
  \{D_{\rm matter}[M],D_{\rm matter}[N]\}&=D_{\rm matter}[MN'-NM'] \label{DDm}\\
  \{H_{\rm matter}[M],D_{\rm T}[N]\}&=-H_{\rm matter}[NM'] \label{HDm}\\
  \{H_{\rm matter}[M],H_{\rm matter}[N]\}&=D_{\rm
    matter}\left[\bar{\beta}|E^x|(E^{\varphi})^{-2}(MN'-NM')\right] \label{HHm}
\end{align}
where $D_{\rm T}[N]:=D[N]+D_{\rm matter}[N]$ is the total diffeomorphism
constraint, including the gravitational part.  Classically one would have
$\bar{\beta}=1$ (and $\mathcal{H}_{\rm matter}$ would only depend on the triad
fields), but here we are allowing for a correction function
$\bar{\beta}(K_{\varphi},E^x)$ to take into account possible deformations of
the matter part as in (\ref{HHbeta}). Therefore, to compute brackets we assume
that $\mathcal{H}_{\rm matter}$ may also depend on $K_{\varphi}$ (but not on
$K_x$, nor on derivatives of $K_{\varphi}$ or the triad). The brackets of
total Hamiltonian constraints, combining gravity and matter contributions,
then do not decouple from each other, and cross-terms will have to be
considered below. (Cross-terms must vanish in this case according to the
argument given at the beginning of this subsection, but will do so only with
additional restrictions on the modification functions.)

Examples of such deformed matter systems include the scalar field,
dust and null dust: In the first case, we have a canonical pair
\begin{equation}
\{\phi(x),P_\phi(y)\}=\frac{1}{4\pi}\delta(x,y),
\end{equation}
and corresponding constraints
\begin{equation}
D_{\rm matter}[N]=4\pi\int{\rm d} x\,NP_\phi\,\phi' \,,
\end{equation}
\begin{equation}
H_{\rm matter}[M]=4\pi\int{\rm d}
x M\left(\nu\frac{P_\phi^2}{2|E^x|^{1/2}E^{\varphi}}
+\sigma\frac{|E^x|^{3/2}\phi'\,^2}{2E^{\varphi}}
+|E^x|^{1/2}E^{\varphi}\frac{U(\phi)}{2}\right),
\end{equation}
with correction functions $\nu(K_{\varphi},E^x)$ and $\sigma(K_{\varphi},E^x)$
such that $\bar{\beta}=\nu\sigma$. For dust fields \cite{BrownKuchar}, we have
two canonical pairs with
\begin{equation} \label{dustCPairs}
\{\tau(x),P_\tau(y)\}=\{\Phi(x),P_\Phi(y)\}=\frac{1}{4\pi}\delta(x,y)\,,
\end{equation}
and a contribution
\begin{equation}
D_{\rm matter}[N]=4\pi\int {\rm d}x\, N\left(P_\tau\tau'+P_\Phi\Phi'\right)
\end{equation}
to the diffeomorphism constraint, while the matter part of the Hamiltonian
constraint is
\begin{equation}
H_{\rm matter}[M]=4\pi\int {\rm d}x\,
M\sqrt{P_\tau^2+\bar{\beta}\frac{|E^x|}{(E^\varphi)^2}(P_\tau
  \tau'+P_\Phi\Phi')^2}  \,.
\end{equation}
For null dust fields \cite{NullDust}, only the the second canonical pair in
(\ref{dustCPairs}) survives and
\begin{equation}
H_{\rm matter}[M]=4\pi\int {\rm d}x\,
M\sqrt{|\bar{\beta}|}\frac{\sqrt{|E^x|}}{E^\varphi}|P_\Phi\Phi'| \,.
\end{equation}

Starting from the classical linear combination of constraints
\[
\tilde{\mathcal{C}}_{\rm
  T}=\frac{(E^x)'}{E^{\varphi}}(\mathcal{H}+\mathcal{H}_{\rm matter})-
2\frac{K_{\varphi}\sqrt{|E^x|}}{E^{\varphi}}(\mathcal{D}+\mathcal{D}_{\rm matter})\,,
\]
one may replace $K_{\varphi}^2$, $K_{\varphi}K_{\varphi}'$ and $K_{\varphi}$
multiplying the matter part of the diffeomorphism constraint by three
different functions $f_1$, $F_2$ and $F_{\rm matter}$. We therefore define
\begin{equation}  \label{Cmatter}
\tilde{\mathcal{C}}_{\rm matter}:=\frac{(E^x)'}{E^{\varphi}}\mathcal{H}_{\rm matter}-
2\frac{F_{\rm matter}(K_{\varphi},E^x)\sqrt{|E^x|}}{E^{\varphi}}\mathcal{D}_{\rm matter}
\end{equation}
and
\begin{equation}
\tilde{\mathcal{C}}_{\rm T}:=\tilde{\mathcal{C}}+\tilde{\mathcal{C}}_{\rm matter}\,.
\end{equation}
In the first term of gravitational contributions to the constraint, we now
consider a more-general
modified expression:
\begin{align}
\tilde{C}[M]=&-\frac{1}{2G}\int{\rm d}
x\,M\bigg(|E^x|^{-1/2}(E^x)'(1+f_1(K_{\varphi},E^x))+
2|E^x|^{1/2}F_2(K_{\varphi},K_{\varphi}',E^x)
\nonumber\\
&-\frac{(E^x)'}{4(E^{\varphi})^2}\left(4|E^x|^{1/2}(E^x)''+
|E^x|^{-1/2}((E^x)')^2\right)+
\frac{|E^x|^{1/2}((E^x)')^2(E^{\varphi})'}{(E^{\varphi})^3}\bigg)
\end{align}
where $K_{\varphi}^2$ has been replaced by a function $f_1$ of $K_{\varphi}$
(and possibly $E^x$), and $K_{\varphi}K_{\varphi}'$ by a function $F_2$ of
these same variables. We will also assume the orientation $E^x>0$.

Using the equivalent expression
\begin{equation} \label{tildeC2}
\tilde{C}[M]=-\frac{1}{2G}\int{\rm d} x\,M\left\{\frac{{\rm d}}{{\rm d}x}\left[
    2\,|E^x|^{1/2}
 \left(1- \Gamma_{\varphi}^2\right)\right]
+2(|E^x|^{1/2})'f_1+2|E^x|^{1/2}F_2\right\}\,,
\end{equation}
it is straight-forward to see that requiring the term inside the parenthesis
to be a total derivative restricts $f_1$ and $F_2$ to be independent of $E^x$,
and $F_2$ to be linear in $K_{\varphi}'$:
\begin{equation} \label{f2f1_2}
F_2(K_{\varphi},K_{\varphi}')=2f_2(K_{\varphi})K_{\varphi}'  \qquad
\text{with} \qquad  2f_2=\frac{{\rm d}
  f_1}{{\rm d}K_{\varphi}}\,.
\end{equation}
(Substituting the first condition back in (\ref{tildeC2}) we recover
(\ref{tildeC1}) and the second condition is again the same as the one obtained
from effective models for the closure of the modified Hamiltonian and
diffeomorphism constraints.)

Using (\ref{DDm}), (\ref{HDm}) and (\ref{HHm}), we compute the bracket
\begin{align}
\{\tilde{C}_{\rm T}[M],\tilde{C}_{\rm T}[N]\}=&\,\{\tilde{C}[M],\tilde{C}[N]\}
% \notag\\
\,+\int{\rm d} x(MN'-NM')\nonumber\\
&\times\bigg\{
\frac{|E^x|}{(E^{\varphi})^2}\left[\frac{((E^x)')^2}{(E^{\varphi})^2}
\left(\bar{\beta}-\frac{\partial
      F_{\rm matter}}{\partial K_{\varphi}}\right)+2F_{\rm matter}\left(2F_{\rm
      matter}-\frac{\partial
      F_2}{\partial K_{\varphi}'}\right)\right]\mathcal{D}_{\rm matter}
\nonumber\\
&-\frac{|E^x|^{1/2}(E^x)'}{(E^{\varphi})^2}\left(2F_{\rm matter}-\frac{\partial
    F_2}{\partial K_{\varphi}'}\right)\left(\mathcal{H}_{\rm matter}-E^{\varphi}\,
\frac{\partial\mathcal{H}_{\rm matter}}{\partial E^{\varphi}}\right)
\nonumber\\
&+\frac{|E^x|^{1/2}((E^x)')^3}{2(E^{\varphi})^4}\,
\frac{\partial\mathcal{H}_{\rm matter}}{\partial K_{\varphi}}
\, \bigg\} \,. \label{CTCT}
\end{align}
(For details, see App.~\ref{a:CC2}.)  We first note that the bracket
\begin{align}
\{\tilde{C}[M],\tilde{C}[N]\}=&-\frac{1}{2G}\int{\rm d}
x\,(MN'-NM')\frac{|E^x|\,((E^x)')^2}{(E^{\varphi})^3}
\bigg[\frac{\partial F_2}{\partial K_{\varphi}}-\frac{\partial^2
  F_2}{\partial K_{\varphi}\partial K_{\varphi}'}K_{\varphi}'  \nonumber\\
&+\left(\frac{1}{2|E^x|}\left(\frac{\partial f_1}{\partial
      K_{\varphi}}-\frac{\partial
      F_2}{\partial K_{\varphi}'}\right)-\frac{\partial^2F_2}{\partial
    E^x \partial
    K_{\varphi}'}\right)(E^x)'
-\frac{\partial^2F_2}{(\partial
  K_{\varphi}')^2}\,K_{\varphi}''\,\bigg]\,  \label{CCvacBracket}
\end{align}
by itself may form a closed system only if it vanishes identically: since
(\ref{CCvacBracket}) does not depend on $K_x$ and $(E^x)''$,
$\{\tilde{C}[M],\tilde{C}[N]\}=
\mathcal{F}_{\tilde{\mathcal{C}}}\tilde{\mathcal{C}}+
\mathcal{F}_{\mathcal{D}}\mathcal{D}$ implies
$\mathcal{F}_{\tilde{\mathcal{C}}}=\mathcal{F}_{\mathcal{D}}=0$. This is the
Abelianization condition in the vacuum case.

The vanishing of the $K_{\varphi}''$ term again implies that $F_2$ must depend
linearly on $K_{\varphi}'$.  Using this condition, all terms proportional to
$K_{\varphi}'$ cancel out. The vanishing of the first term and the
$(E^x)'$-term imply that $F_2$ has the form
$F_2=2f_2(K_{\varphi},E^x)K_{\varphi}'+f_3(E^x)$, for a general function $f_3$
of the triad component $E^x$, as well as
\begin{equation}
\frac{\partial f_1}{\partial K_{\varphi}}-2f_2=4|E^x|\frac{\partial
  f_2}{\partial E^x}\,.
\end{equation}
This requirement matches (\ref{f2f1_2}) in the case of correction functions
independent of $E^x$.

With these conditions, we can now look at the additional contributions to
(\ref{CTCT}) in the presence of matter.  Using the expression obtained for
$F_2$ and requiring the total bracket (\ref{CTCT}) to be zero, we must have
\begin{equation}  \label{abelianizationCond2}
F_{\rm matter}=f_2\quad,\quad
\bar{\beta}=\frac{\partial f_2}{\partial K_{\varphi}}\quad\mbox{and}\quad
\frac{\partial\mathcal{H}_{\rm matter}}{\partial K_{\varphi}}=0\,.
\end{equation}
The last condition in (\ref{abelianizationCond2}) tells us that no deformation
of the matter part depending on $K_{\varphi}$ is consistent with
Abelianization (or a closed system).  Furthermore, in the case of deformations
of the matter Hamiltonian independent of curvature, $\bar{\beta}(E^x)$ can
only be a function of the triad. Thus also in this case, the second condition
in (\ref{abelianizationCond2}) implies that the only possible dependence on
$K_{\varphi}$ of the whole system is the classical one.  If a deformation
consistent with Abelianization exists, it must contain other derivatives of
the fields. Remarkably, however, in the classical case with $\bar{\beta}=1$
Abelianization of the constraint $\tilde{C}_{\rm T}[M]$ follows for general
matter systems satisfying (\ref{DDm})--(\ref{HHm}), not just for a scalar
field.

\subsubsection{Maxwell field}
\label{s:Maxwell}

To arrive at the negative conclusions above, it was crucial that the matter
contribution to the diffeomorphism constraint is assumed to be non-zero.  It
is well-known, however, that substiting a spherically symmetric ansatz in the
canonical action for a Maxwell field leads to a consistent reduced system with
a vanishing contribution to the diffeomorphism (or vector) constraint
\cite{SphSymmHam}.  This property leaves the possibility open for a consistent
Abelianization of the Einstein-Maxwell system (which, however, does not have
local degrees of freedom in spherical symmetry).

Indeed, in this case the canical pairs are
\begin{equation}
\{\mathcal{A}_x(x),P(y)\}=\frac{1}{4\pi}\delta(x,y)\,,
\end{equation}
with $\mathcal{A}_x(x)$ the sole spatial radial component of the vector
potential and $\mathcal{E}^x:=P\sin\vartheta$ the only non-zero radial
component of the (densitized) electric field. The contribution to the
Hamiltonian constraint is
\begin{equation}
H_{\rm matter}[M]=4\pi\int {\rm d}x\,M\frac{E^\varphi P^2}{2\,|E^x|^{{3}/{2}}}\,,
\end{equation}
and there is the additional (Maxwell) Gauss constraint:
\begin{equation}
{G}_{\rm Maxwell}[\Lambda]=4\pi\int {\rm d}x\,\Lambda\,P'\,.
\end{equation}
There is no contribution to the vector constraint obtained from the
$\{H,H\}$-bracket, so the system does not satisfy (\ref{DDm})--(\ref{HHm}) but
instead
\begin{equation}
  D_{\rm T}[N]=D[N]\,,
\end{equation}
\begin{equation}
  \{H_{\rm matter}[M],D_{\rm T}[N]\}=-H_{\rm matter}[NM']-G_{\rm
    Maxwell}\big[MNE^\varphi|E^x|^{-{3}/{2}}P\big]\,,
\end{equation}
\begin{equation}
  \{H_{\rm matter}[M],H_{\rm matter}[N]\}=0\,.
\end{equation}

As before, one may also consider a deformed system (which satisfies the same
bracket relations) with
\begin{equation}
H_{\rm matter}[M]=4\pi\int {\rm d}x\,M\frac{\nu\, E^\varphi
  P^2}{2\,|E^x|^{{3}/{2}}}\,,
\end{equation}
and correction function $\nu(K_\varphi,E^x)$.  The combined constraint
$\tilde{C}_{\rm T}[M]$ results from taking $\mathcal{D}_{\rm matter}=0$ in
expression (\ref{Cmatter}), so that now $\{\tilde{C}_{\rm
  matter}[M],\tilde{C}_{\rm matter}[N]\}=0$ and only the last term in
(\ref{CTCT}) survives:
\begin{align}
\{\tilde{C}_{\rm T}[M],\tilde{C}_{\rm T}[N]\}=&\,\{\tilde{C}[M],\tilde{C}[N]\}
% \notag\\
\,+\int{\rm d} x(MN'-NM')\frac{|E^x|^{1/2}((E^x)')^3}{2(E^{\varphi})^4}\,
\frac{\partial\mathcal{H}_{\rm matter}}{\partial K_{\varphi}}
\,.
\end{align}
Thus, a consistent Abelian deformation is always possible, but again only as
long as the correction function $\nu$ is independent of curvature. The
gravitational contribution to the Hamiltonian constraint, however, can be
modified in a curvature-dependent way. Nevertheless, this model is not a
counter-example to our statements that no covariant holonomy-modified models
with local degrees of freedom are known, because there are no local degrees of
freedom in the spherically symmetric Einstein--Maxwell system. As we shall
recall in the next section, these properties are again fully compatible with
results \cite{RNDeformed} using effective methods and the requirement of
anomaly-freedom.

We can interpret this system as further circumstantial evidence that local
degrees of freedom seem to be responsible for making it more difficult (if not
impossible) to find covariant models with holonomy modifications. The
spherically symmetric Einstein--Maxwell system can obey the required
consistency conditions, but only because the Gauss constraint allows one to
eliminate the new kinematical degree of freedom, given by the Maxwell fields,
from the diffeomorphism constraint. The same constraint, $P'=0$ in its local
version, removes the new kinematical degree of freedom from the reduced phase
space. In contrast to the scalar or dust examples, the non-gravitational local
kinematical degree of freedom therefore does not lead to local physical
degrees of freedom, which then do not seem to present an obstacle to a
consistent holonomy-modified model.

Looking back at these calculations, the modified Einstein--Maxwell system can
be consistent despite the fact that there is no contribution to the
diffeomorphism constraint because in this case the matter contribution
$\int{\rm d}x\,\delta N\mathcal{A}_xP'$ to the infinitesimal generator of
radial diffeomorphisms is a multiple of the Gauss constraint. Again, this is a
special property of reduced models and unlikely to extend to general
configurations. One may consistently define the Einstein--Maxwell constrained
system as in \cite{RNDeformed}, with non-zero contribution
\begin{equation} \label{DiffMax}
D_{\rm matter}[N]=-4\pi\int{\rm d}x\,N\mathcal{A}_xP'\,.
\end{equation}
However, this alternative set of constraints satisfies
(\ref{DDm})--(\ref{HHm}) with $\bar{\beta}=0$ and hence does not lead to an
Abelian deformation. (In fact, even classically, the corresponding system of
constraints $C_{\rm T}[M]$ and $D_{\rm T}[N]$ is not closed unless the Gauss
constraint is included.) Even though the two initial systems of constraints
with different contributions to $D_{\rm T}[N]$ are equivalent, the two systems
derived from them by substituting the Hamiltonian constraint with $C_{\rm
  T}[M]$ are not.

The generator of spatial diffeomorphisms, (\ref{DiffMax}), has also been used
in \cite{LoopReiss} in the context of Abelianization. While Abelianization of
normal hypersurface deformations could be achieved in this case, it was
possible only by fixing the U(1)-gauge of the Maxwell contribution. Our
construction leads to a more general result, showing Abelianization even if no
partial gauge fixing is used.

\subsection{Impossible modifications}

We will now verify explicitly that the impossibility of obtaining a
(partially) Abelian algebra from deformations of the classical
$\tilde{\mathcal{C}}_{\rm T}$ constraint is consistent with negative results
for an anomaly-free deformed constraint algebra.

Again, consider a classical spherically symmetric matter system with
non-derivative couplings, and such that correction functions in the Hamiltonian
$H_{\rm T}[M]=H[M]+H_{\rm matter}[M]$ do not contain derivatives of the
gravitational fields $K$ and $E$. It is easy to see that if the deformed
vacuum algebra satisfies
\begin{equation}
\{H[M],H[N]\}=D[{\beta}|E^x|(E^{\varphi})^{-2}(MN'-NM')]\,
\end{equation}
with a correction function ${\beta}$ depending on the connection or extrinsic
curvature, and the matter contribution to the diffeomorphism constraint is
non-trivial, then the matter Hamiltonian must be deformed with correction
functions also depending on extrinsic curvature. (This is at least true if we
assume no second or higher derivatives of the matter fields.) Indeed, if we
assume $\mathcal{H}_{\rm matter}$ to be independent of $K_x$ and $K_{\varphi}$ it
follows that
\[
S:=\{H[M],H_{\rm matter}[N]\}\,-\,(M\leftrightarrow N) =0\,,
\]
and therefore the `crossed' or `mixed' brackets vanish and we have
\[
\{H_{\rm T}[M],H_{\rm T}[N]\}=\{H[M],H[N]\}+\{H_{\rm matter}[M],H_{\rm matter}[N]\}\,.
\]
For a first-class algebra we must have
\[
\{H_{\rm matter}[M],H_{\rm matter}[N]\}=D_{\rm
  matter}[{\beta}|E^x|(E^{\varphi})^{-2}(MN'-NM')] \,.
\]
There cannot be additional multiples of the (total) Hamiltonian constraint
since the latter contains second derivatives of $E^x$.  However, the
right-hand side of the above expression for the bracket depends on curvature,
so the left hand side, that is $H_{\rm matter}[M]$, must depend on curvature
after all.

Motivated by the previous observations and by consistent deformations with
inverse-triad corrections obtained in \cite{JR,LTBII,ModCollapse}, we will
consider matter systems which additionally satisfy (\ref{DDm}), (\ref{HDm})
and (\ref{HHm}) with a correction function $\bar{\beta}(K_x,K_{\varphi},E^x)$
depending on both extrinsic curvature components and $E^x$. (The scalar and
dust fields above with deformation functions also depending on $K_x$ satisfy
these conditions.)  For these systems or any other model with matter
Hamiltonians depending on connection or extrinsic-curvature components, we
have
\begin{align}
S=&\int{\rm d} x\,(MN'-NM')\left[\left(
\frac{|E^x|^{-\frac{1}{2}}(E^x)'}{2E^{\varphi}}-
\frac{|E^x|^{\frac{1}{2}}(E^{\varphi})'}{(E^{\varphi})^2}\right)
\frac{\partial\mathcal{H}_{\rm matter}}{\partial K_x}
+\frac{|E^x|^{\frac{1}{2}}(E^x)'}{2(E^{\varphi})^2}
\frac{\partial\mathcal{H}_{\rm matter}}{\partial K_{\varphi}}
\right] \nonumber\\
&+\int{\rm d} x\,(MN''-NM'')\frac{|E^x|^{\frac{1}{2}}}{E^{\varphi}}
\frac{\partial\mathcal{H}_{\rm matter}}{\partial K_x}\,.
\end{align}
As shown in \cite{HigherSpatial}, variations by $MN'-NM'$ and $MN''-NM''$ are
independent, so that $\partial{\cal H}_{\rm matter}/\partial K_x=0$.
Therefore, restricting now to $K_x$-independent corrections,
\begin{align}
\{H_{\rm T}[M],H_{\rm T}[N]\}=&\int{\rm d} x\,(MN'-NM')\bigg(
\frac{|E^x|}{(E^{\varphi})^2}
\left({\beta}\mathcal{D}+\bar{\beta}\mathcal{D}_{\rm matter}\right)%%\notag\\
&+\frac{|E^x|^{\frac{1}{2}}(E^x)'}{2(E^{\varphi})^2}
\frac{\partial\mathcal{H}_{\rm matter}}{\partial
  K_{\varphi}}\bigg)\,.   \label{HTHT}
\end{align}

It is now easy to see that the last term cannot be a linear combination
containing the total Hamiltonian because the latter contains second-order
derivatives of $E^x$ in its gravitational part while the former does
not. Since $\mathcal{H}_{\rm matter}$ must be independent of $K_x$, this last
term cannot contain a multiple of the gravitational part of the diffeomorphism
constraint $\mathcal{D}$ either.  Hence the only possibilities left for a
closed algebra are that the last term vanishes or that it is a multiple of
$\mathcal{D}_{\rm matter}$. Since we are assuming $\mathcal{D}_{\rm
  matter}\neq 0$, this last possibility is, however, inconsistent since it
would require ${\beta}$ to depend on $(E^x)'$.  It then follows again
  that deformations of the matter Hamiltonian must be independent of
  $K_{\varphi}$:
\[
\frac{\partial\mathcal{H}_{\rm matter}}{\partial K_{\varphi}}=0
\quad\mbox{and}\quad
\beta=\bar{\beta}\,.
\]
If $\mathcal{H}_{\rm matter}$ is independent of both curvature components, then
$\bar{\beta}$ is necessarily independent of curvature too and the last
condition above precludes any vacuum deformation such as (\ref{betaK})
depending on curvature.  We have come full circle, in this case the only
consistent deformations of the combined gravity-matter system independent of
$K_x$ must also be independent of $K_{\varphi}$. Only triad-dependent
deformations of the type found in \cite{JR,LTBII,ModCollapse} are allowed.

For Maxwell fields, there is no contribution from the $\{H_{\rm
  matter}[M],H_{\rm matter}[N]\}$-bracket, and therefore
$\bar{\beta}\mathcal{D}_{\rm matter}=0$ in (\ref{HTHT}). Deformations of the
gravitational and matter parts of the Hamiltonian effectively `decouple' and
we see that consistent or anomaly-free deformations are possible with
$K_\varphi$-dependent deformations (\ref{betaK}) of the gravitational part and
an undeformed or deformed but curvature-independent matter Hamiltonian.

\section{Conclusions}

Abelianization of normal hypersurface deformations can eliminate structure
functions from constraint brackets and thereby open up access to standard
quantization methods applied to gravitational models. However, by itself, this
result leaves the question of covariance unaddressed, which is important for
gravitational theories. As shown here, covariance of modified theories is
indeed non-trivial in this setting, and it is not always realized: A standard
holonomy modification of the Abelianized constraint does not lead to
hypersurface-deformation generators with the correct classical limit if a
scalar field or other matter with local physical degrees of freedom are
coupled to gravity.

In our general discussion of covariance in canonical systems, we have
highlighted the important distinction between background treatments and
background-independent theories. Even if back-reaction is not considered,
there is a difference between these two cases as regards covariance in
non-classical systems. Hypersurface-deformation generators may then be
deformed in different ways as one departs from the classical limit, but a
consistent gravity-matter system requires the same deformation of both
ingredients. A background treatment in which covariance is required separately
for gravity and matter, on the other hand, may formally give rise to more
options. As an example, the holonomy-modified scalar model of
\cite{LoopSchwarz2} does not correspond to a covariant gravity-matter system,
as shown here, but the actual constructions of \cite{LoopSchwarz2} make use of
a gravitational background and may be formally consistent. (We note that two
different kinds of modifications appear in \cite{LoopSchwarz2}, holonomy
modifications and a discretization of the scalar Hamiltonian. While the latter
is in the foreground in \cite{LoopSchwarz2}, we have tested only the former in
the present paper. Covariance conditions on discretized scalar Hamiltonians
remain to be explored, but possible discrete versions of
hypersurface-deformation brackets are known \cite{DiscDirac}.)

An interesting result is also the fact that there seems to be complete
agreement on this question, addressed with different methods: Abelianization
and anomaly freedom implemented with effective techniques as introduced in
\cite{ConstraintAlgebra} in the context of cosmological perturbations.  This
convergence of results obtained by different methods gives further support to
the phenomenon of signature change discovered by an analysis of canonical
effective models \cite{Action}. At first sight, it might seem that the
constructions of \cite{LoopSchwarz,LoopSchwarz2} do not lead to modified
space-time structures in spherically symmetric models, unlike what effective
calculations have shown in the same models \cite{JR,HigherSpatial}. However,
if one actually poses the question of covariance and space-time structure in
the constructions of \cite{LoopSchwarz,LoopSchwarz2}, one finds, as shown
here, that covariance requires the Hamiltonian constraints to be modified with
the same restriction (\ref{f2f1}) as found in \cite{JR,HigherSpatial} for
anomaly-free effective models. If $K_{\varphi}^2$ is replaced by some function
$f(K_{\varphi})$, in effective and Abelianized models the same modified
brackets
\begin{equation}
 \{H[N_1],H[N_2]\} = -D[\beta(K_{\varphi})(|E^x|/(E^{\varphi})^2) (N_1N_2'-N_2N_1')]
\end{equation}
are realized for generators of hypersurface deformations,
with a non-trivial function
\begin{equation} \label{beta}
 \beta(K_{\varphi}) = \frac{1}{2}\frac{\partial^2 f}{\partial K_{\varphi}^2}\,.
\end{equation}
(Signature change is indicated by $\beta$ changing sign, which always happens
if $f(K_{\varphi})$ has a local maximum. For the popular modification
$f(K_{\varphi})=\delta^{-2}\sin^2(\delta K_{\varphi})$, for instance,
$\beta(K_{\varphi})=\cos(2\delta K_{\varphi})$.)  The agreement of results is
promising, but at the same time one then has to take seriously the resulting
modified space-time structures at high curvature, which can lead to problems
of indeterminism and Cauchy horizons for black holes \cite{Loss} or global
issues for cosmological perturbation equations \cite{SigImpl}.

In this light, the language used in \cite{LoopSchwarz,LoopSchwarz2}, speaking
about quantum systems {\em on quantum space-time} does not seem justified
because covariance conditions, which are usually understood as being crucial
for space-time theories, have not been checked. (This language goes back to
cosmological constructions in \cite{QFTCosmo,AAN}, where it seems equally
unjustified because the background minisuperspace models used in these
examples do not even allow one to test covariance and the consistency of
quantum space-time structures. Instead, metric structures are merely
postulated.)  In the scalar model, no consistent space-time structure of the
holonomy-modified theory is known, so that it seems unclear how to use formal
solutions of these systems for an analysis of Hawking radiation, the stated
aim of \cite{LoopSchwarz2}.

At present, it is not known whether covariance can always be realized in the
presence of holonomy modifications from loop quantum gravity, even if one
restricts oneself to the rather tractable spherically symmetric
models. Especially the presence of local physical degrees of freedom seems to
pose a challenge, as indicated by the general matter models considered here
(as well as the spherically symmetric Einstein--Maxwell system as discussed in
Sec.~\ref{s:Maxwell}) and the polarized Gowdy models of \cite{GowdyCov}. This
result of our paper might pose a challenge to loop quantum gravity. We
certainly did not discuss full quantizations of the models considered, but if
the theory is to have the correct semiclassical limit, brackets of the form
analyzed here will be encountered in some way.

The partial nature of our no-go results can be used to suggest how covariant
holonomy-modified models with local degrees of freedom could possibly be
realized. One way to avoid the negative conclusions would be to include higher
spatial derivatives of the matter field. Such terms are expected in continuum
effective models of loop quantum gravity because matter fields and their
standard derivative terms in the Hamiltonian have to be discretized for an
operator acting on spin-network states \cite{QSDV}. For anomalies to cancel
out, holonomy modifications in the gravitational contribution to the
constraint would have to be carefully adjusted to matter discretizations. So
far, these two quantization steps have been considered as independent, but
off-shell anomaly-freedom may force one to combine them. If a consistent
version then becomes possible, it would have several unexpected features, in
addition to making consistent models rather tightly constrained. First, for
the covariance conditions of the gravitational background and the matter
system to match, the matter discretization would have to depend on extrinsic
curvature because the modified structure function (\ref{beta}) of a covariant
holonomy-modified background has such a dependence. Secondly, holonomy
modifications in one direction (here, an angular direction in spherically
symmetric models which gives rise to point holonomies of $K_{\varphi}$) would
have to be closely related to the matter discretization in another direction
(here, the radial one so as to have higher spatial derivatives). It is not
clear whether covariant models can be found by implementing these features,
evading our no-go results. (For radial holonomy modifications in vacuum
spherically symmetric models, higher spatial derivatives do not seem to help
much \cite{HigherSpatial}.) Nevertheless, there is a chance that it would be
fruitful to match covariance conditions of gravitational terms with holonomy
modifications, as studied in \cite{JR,HigherSpatial} and in the present paper,
with methods to obtain consistent discretizations as studied for instance in
\cite{ConsistDisc,UniformDisc,DiscDirac}.

\section*{Acknowledgements}

We are grateful to Rodolfo Gambini and Jorge Pullin for discussions.  This
work was supported in part by NSF grant PHY-1307408.
JDR was supported by CONACYT 0177840 grant.

\begin{appendix}

\section{Diffeomorphism constraint}
\label{a:Diff}

The diffeomorphism constraint in loop quantum gravity is usually not
constructed by writing the classical expression in terms of holonomies and
inserting basic operators, but rather by lifting the spatial flow generated by
the constraint to the state space \cite{ALMMT}. (See \cite{DiffeoOp} for an
alternative.)

In spherically symmetric models \cite{SphSymm}, one can represent states by
referring to an orthonormal basis
\begin{equation} \label{states}
 \psi_{\{(x_1,k_1,\mu_1),\ldots,(x_n,k_n,\mu_n)\}}[K_x,K_{\varphi}] =
   \prod_{j=1}^n \exp\left(ik_j \smallint_{x_j}^{x_j+1} K_x{\rm d}x\right)
\exp\left(i\mu_j K_{\varphi}(x_j)\right)
\end{equation}
with integer $k_j$, real numbers $\mu_j$, and $x_j$ in the radial
manifold. (For simplicity, we assume the radial manifold to be compact. In the
notation used to write states, we set $k_{n+1}=0$.) Spatial
diffeomorphisms $\Phi$ can easily be represented unitarily by
\begin{equation} \label{Phi}
 \hat{\Phi} \psi_{\{(x_1,k_1,\mu_1),\ldots,(x_n,k_n,\mu_n)\}}:=
\psi_{\{(\Phi(x_1),k_1,\mu_1),\ldots,(\Phi(x_n),k_n,\mu_n)\}}\,.
\end{equation}
This action can be used to factor out spatial diffeomorphisms by group
averaging, but it does not define a diffeomorphism constraint: States with
different $\{x_1,\ldots,x_n\}$ are orthogonal to each other, so that one
cannot take a $t$-derivative of the quantized flow of a 1-parameter family
$\Phi_t=\exp(t v)$ with a spatial vector field $v$ as an infinitesimal
generator.

\subsection{Effective constraints}

In a continuum effective theory, on the other hand, there should be a
well-defined version of the diffeomorphism constraint, possibly with quantum
corrections. For instance, in the canonical framework of
\cite{EffAc,EffCons,EffConsRel}, the effective constraint would be computed as
the expectation value of $\hat{\Phi}$ in a suitable class of semiclassical
states obtained by superpositions of the basis states. For a local effective
theory (and therefore the classical limit) to exist, these superposed states
must be such that expectation values $\langle \hat{E}^x\rangle$ and so on are
differentiable functions of $x$ in some coarse-graining approximation. A
derivative expansion of these or more-complicated expectation values (such as
the Hamiltonian constraint) then gives rise to a theory with gauge
transformations of infinitesimal diffeomorphisms acting on effective fields.

In order to compute an effective constraint, one need not construct explicit
semiclassical states, which would be challenging in models of loop quantum
gravity. Instead, one parameterizes states by expectation values and moments
of basic operators, so that a semiclassical regime can be specified more
easily by a certain hierarchy of the moments by powers of $\hbar$.  By the
same condition, the derivative expansion can be combined with a semiclassical
expansion, in which the classical diffeomorphism constraint is extended by
moment terms. Not only expectation values of the basic operators but also
their fluctuations and higher moments are then subject to gauge
transformations.

In addition to expectation values of basic operators quantizing $E^x$, $K_x$,
$E^{\varphi}$ and $K_{\varphi}$ in the case of spherically symmetric models,
the moments are defined as
\begin{eqnarray}
\Delta\left[\left(E^{\varphi}\right)^{n_1}\left(E^{x}\right)^{n_2}
\left(K_{\varphi}\right)^{n_3}\left(K_{x}\right)^{n_4}\right]:=
\left\langle \left(\widehat{\Delta E^{\varphi}}\right)^{n_1}
  \left(\widehat{\Delta E^{x}}\right)^{n_2} \left(\widehat{\Delta
      K_{\varphi}}\right)^{n_3} \left(\widehat{\Delta
      K_{x}}\right)^{n_4}\right\rangle_{\rm symm}
\end{eqnarray}
in totally symmetric ordering, where $\widehat{\Delta \zeta}:=\hat\zeta -
\langle\hat\zeta\rangle$ if $\zeta$ represents a generic phase space
variable. In a loop quantization, one would use holonomy operators instead of
quantized components of extrinsic curvature.  These variables form a phase
space, with a Poisson bracket based on an extension of
\begin{equation} \label{Poisson}
 \{\langle\hat{A}\rangle,\langle\hat{B}\rangle\} =
 \frac{\langle[\hat{A},\hat{B}]\rangle}{i\hbar}
\end{equation}
to moments by the Leibnitz rule.

For expectation values of basic operators, (\ref{Poisson}) reduces to the
classical bracket.  The bracket (\ref{Poisson}) applied to moments is not the
only extension of the classical bracket one could think of, but it is
distinguished by the fact that a closed commutator algebra of some set of
operators, such as some first-class constraint operators $\hat{C}_I$, implies
a closed algebra of effective constraints, defined as $\langle\widehat{\rm
  pol}\hat{C}_I\rangle$ with polynomials $\widehat{\rm pol}$ in basic
operators, under Poisson brackets. One can therefore analyze the possibility
of first-class quantizations by computing Poisson brackets of effective
constraints, which in most cases is much more feasible than analyzing the
possibility of closed commutators. The effective constraints can be computed
in terms of the moments by Taylor expanding the expectation value in
$\langle\widehat{\Delta\zeta}\rangle$ \cite{EffAc,EffCons,EffConsRel}.

In models with local kinematical degrees of freedom, we proceed formally in
order to illustrate the main features. (But see \cite{CW} for a demonstration
that canonical effective methods can also be applied to quantum field
theories.) For the diffeomorphism constraint of the spherically symmetric
vacuum model, given in (\ref{Diffeoclassical}), we have an infinite family of
effective constraints $D[N]_{\text{pol}}:=
\left\langle\widehat{\text{pol}}\,\hat{D}[N]\right\rangle$ where
$\widehat{\text{pol}}$ now stands for arbitrary polynomials in the
$\widehat{\Delta \zeta}$ of spherically symmetric variables. We assume that we
have selected a consistent factor-ordering choice for the operator
$\hat{D}[N]$, which in this case is known to exist \cite{SphSymmOp}.  For
semi-classical states, we have
\begin{eqnarray}
\Delta\left[\left(E^{\varphi}\right)^{n_1}\left(E^{x}\right)^{n_2}
\left(K_{\varphi}\right)^{n_3}\left(K_{x}\right)^{n_4}\right] \equiv
\mathcal{O}\left(\hbar^{(n_1+n_2+n_3+n_4)}\right)\,.
\end{eqnarray}
This hierarchy allows us to consider a closed system of finitely many local
effective constraints to any fixed order in $\hbar$, after expanding each of
these constraints (starting with the diffeomorphism constraint for
$\text{pol}=1$) in terms of basic expectation values
$\langle\hat{\zeta}\rangle$ and the moments.

As follows from general considerations of effective constrained systems
\cite{EffCons,EffConsRel,EffConsQBR}, no new observables arise in this way,
but quantum corrections to the classical reduced phase space appear. For every
new quantum variable given by a moment, there is a higher-order effective
constraint with $\widehat{\rm pol}\not=1$ which fixes the moment or removes it
by the gauge flow. So far, this property has been demonstrated for
finite-dimensional models, but such a result is sufficient for the usual
counting of local degrees of freedom in which one subtracts the number of
constraints plus gauge flows from the number of kinematical degrees of
freedom.

\subsection{Local observables?}

The statement in our last paragraph is in conflict with an observation made in
\cite{LoopSchwarz,LoopSchwarz2}, pointing out a large class of new {\em local}
observables in loop-quantized spherically symmetric models. However, on closer
inspection, these observables have the following, problematic origin: In loop
quantizations such as the one sketched above, one constructs a state space
using auxiliary ingredients in addition to the classical phase-space variables
(or corresponding quantum numbers): While $k_j$ and $\mu_j$ in (\ref{states})
give eigenvalues of the quantized $E^x$ and $E^{\varphi}$, respectively, the
vertex positions $x_j$ have no classical correspondence. By group averaging
(\ref{Phi}), the diffeomorphism constraint is then solved by factoring out the
vertex positions, that is the non-classical ingredients. In the classical
theory, however, the diffeomorphism constraint and its flow provide
non-trivial relationships between $E^x$, $E^{\varphi}$ and their momenta,
which do not follow from the group-averaging construction. By ignoring these
relationships, the loop-quantized theory has additional local observables, but
their meaning is obscure because their origin is the auxiliary vertex
positions introduced for kinematical states. Indeed,
\cite{LoopSchwarz,LoopSchwarz2} explicitly state that their local observables
parameterize the sequence of successive $k_j$, which depends on how the
spurious vertex positions are injected in states. As our discussion of
effective constraints shows, these observables, while they may look like local
degrees of freedom, cannot be part of a local effective theory. And even
though coordinate-dependent vertex positions are averaged over, they leave a
trace in the resulting theory by the missing relationships between kinematical
phase-space variables.

In loop-quantized spherically symmetric models, the implementation of the
diffeomorphism constraint directly follows the full theory \cite{ALMMT}.
Although the diffeomorphism constraint is usually considered well-understood
in loop quantum gravity, several problems of the theory related to its
solutions remain and indicate difficulties both with coordinate independence
(vertex positions affecting observables even after spatial diffeomorphisms
have been factored out) and the classical limit (observables without a place
in local effective theories).

\section{Constraint bracket for matter models}
\label{a:CC2}

We can compute the bracket (\ref{CTCT}) by splitting the gravity and matter
parts and by exploiting the anti-symmetry property:
 \begin{align}
\{\tilde{C}_{\rm T}[M],\tilde{C}_{\rm T}[N]\}=&\{\tilde{C}[M]+
\tilde{C}_{\rm matter}[M],\tilde{C}[N]+\tilde{C}_{\rm matter}[N]\}
\nonumber\\
=&\{\tilde{C}[M],\tilde{C}[N]\}+\{\tilde{C}_{\rm matter}[M],\tilde{C}_{\rm
  matter}[N]\} \nonumber\\
&+\{\tilde{C}[M],\tilde{C}_{\rm matter}[N]\}-\{\tilde{C}[N],\tilde{C}_{\rm
  matter}[M]\}\,.
\end{align}

The gravity part
\[
\{\tilde{C}[M],\tilde{C}[N]\}=2G\int{\rm d}
x\,\frac{1}{2}\frac{\delta\tilde{C}[M]}{\delta K_{\varphi}}
\frac{\delta \tilde{C}[N]}{\delta E^{\varphi}}\,-(M\leftrightarrow
N)
\]
is simple and results in expression (\ref{CCvacBracket}). The `mixed' brackets
are also straight-forward:
\begin{align}
\{\tilde{C}[M],&\tilde{C}_{\rm matter}[N]\}-\{\tilde{C}[N],\tilde{C}_{\rm
  matter}[M]\} \nonumber\\
=&2G\int{\rm d}
x\,\frac{1}{2}\left(\frac{\delta\tilde{C}[M]}{\delta K_{\varphi}}
\frac{\delta\tilde{C}_{\rm matter}[N]}{\delta E^{\varphi}}\,
-\frac{\delta\tilde{C}[M]}{\delta E^{\varphi}}
\frac{\delta\tilde{C}_{\rm matter}[N]}{\delta K_{\varphi}}\right)\,
\,-\,(M\leftrightarrow N)  \nonumber\\
=&\int{\rm d}
x\,(MN'-NM')|E^x|^{1/2}\left(\frac{((E^x)')^2}{2(E^{\varphi})^3}\,
\frac{\partial\tilde{\mathcal{C}}_{\rm matter}}{\partial K_{\varphi}}
\,-\,\frac{\partial
  F_2}{\partial K_{\varphi}'}\,\frac{\partial\tilde{\mathcal{C}}_{\rm
    matter}}{\partial E^{\varphi}}\right)
\nonumber\\
=&\int{\rm d}
x\,(MN'-NM')\frac{|E^x|^{1/2}(E^x)'}{E^{\varphi}}
\left(\frac{((E^x)')^2}{2(E^{\varphi})^3}\,\frac{\partial\mathcal{H}_{\rm
      matter}}{\partial K_{\varphi}}
\,-\,\frac{\partial
  F_2}{\partial K_{\varphi}'}\,\frac{\partial\mathcal{H}_{\rm
    matter}}{\partial E^{\varphi}}\right)
\nonumber\\
&-\frac{|E^x|}{(E^{\varphi})^2}\left(\frac{((E^x)')^2}{(E^{\varphi})^2}\,
\frac{\partial
    F_{\rm matter}}{\partial K_{\varphi}} + 2F_{\rm matter}\,\frac{\partial
    F_2}{\partial K_{\varphi}'}\right){\mathcal{D}}_{\rm matter}
+\frac{|E^x|^{1/2}(E^x)'}{(E^{\varphi})^2}\,\frac{\partial
  F_2}{\partial K_{\varphi}'}\, \mathcal{H}_{\rm matter} \,.
\end{align}

For the matter part we use
 \begin{align}
\{\tilde{C}_{\rm matter}[M],\tilde{C}_{\rm matter}[N]\}=&\{H_{\rm matter}[\tilde{M}]-
D_{\rm matter}[\hat{M}],H_{\rm matter}[\tilde{N}]-D_{\rm matter}[\hat{N}]\}
\nonumber\\
=&\{H_{\rm matter}[\tilde{M}],H_{\rm matter}[\tilde{N}]\}+\{D_{\rm matter}[\hat{M}],D_{\rm matter}[\hat{N}]\}
\nonumber\\
&-\left(\{H_{\rm matter}[\tilde{M}],D_{\rm matter}[\hat{N}]\}-
\{H_{\rm matter}[\tilde{N}],D_{\rm matter}[\hat{M}]\}\right)\,,   \label{HmDmHmDm}
\end{align}
with
\begin{equation}
\tilde{M}:=\frac{(E^x)'}{E^{\varphi}}M\,, \qquad
\hat{M}:=\frac{2F_{\rm matter}\sqrt{|E^x|}}{E^{\varphi}}M
\end{equation}
and similarly for $\tilde{N}$ and $\hat{N}$.  Since $\mathcal{H}_{\rm matter}$
does not depend on $K_x$ and because of anti-symmetry of the bracket we may
use (\ref{HHm}) directly:
\begin{align}
\{H_{\rm matter}[\tilde{M}],H_{\rm matter}[\tilde{N}]\}&=\left.
\{H_{\rm matter}[\tilde{M}],H_{\rm
  matter}[\tilde{N}]\}\right|_{\tilde{M},\hat{N}}
\nonumber\\
&=
D_{\rm matter}[\bar{\beta}|E^x|(E^{\varphi})^{-2}(\tilde{M}\tilde{N}'-\tilde{N}\tilde{M}')]
\nonumber\\
&=D_{\rm matter}[\bar{\beta}|E^x|((E^x)')^2(E^{\varphi})^{-4}(MN'-NM')]\,,
\end{align}
where the notation $|_{\tilde{M},\hat{N}}$ indicates that in the bracket
$\tilde{M}$ and $\hat{N}$ are taken as constant on phase space.  Similarly,
since $\mathcal{D}_{\rm matter}$ does not depend on gravitational variables, we can use
(\ref{DDm}):
\begin{align}
\{D_{\rm matter}[\hat{M}],D_{\rm matter}[\hat{N}]\}&=D_{\rm matter}[\hat{M}\hat{N}'-\hat{N}\hat{N}']
\nonumber\\
&=D_{\rm matter}[4F_{\rm matter}^2|E^x|(E^{\varphi})^{-2}(MN'-NM')].
\end{align}
Computing the last line in (\ref{HmDmHmDm}) is more subtle. First we write
\begin{align}
\{H_{\rm matter}[\tilde{M}],D_{\rm matter}[\hat{N}]\}=\{H_{\rm matter}[\tilde{M}],D_{\rm T}[\hat{N}]\}-
\{H_{\rm matter}[\tilde{M}],D[\hat{N}]\}\,.
\end{align}
One now may check that
\begin{equation}
\{H_{\rm matter}[\tilde{M}],D_{\rm T}[\hat{N}]\}\,-\,(M\leftrightarrow
N)=\left.\{H_{\rm matter}[\tilde{M}],D_{\rm T}[\hat{N}]\}\right|_{\tilde{M},\hat{N}}
\,-\,(M\leftrightarrow
N)\,.
\end{equation}
There are two additional terms (proportional to $MN'-NM'$) arising from the
phase-space dependence of the smearing fields which could add to the bracket:
one coming from the integration by parts of $(E^x)'$ in
$(\delta\tilde{M}/\delta E^x)\mathcal{H}_{\rm matter}\delta D_{\rm
  T}[\hat{N}]/\delta K_x- (M\leftrightarrow N)$ and the other from the
integration by parts of $K_{\varphi}'$ in the gravitational part of the
diffeomorphism constraint in $(\delta H_{\rm matter}[\tilde{M}]/\delta
E^{\varphi})(\delta D_{\rm T}[\hat{N}]/\delta K_{\varphi})-(M\leftrightarrow
N)$. However, these two terms exactly cancel, and hence we may use
(\ref{HDm}):
\begin{align}
\{H_{\rm matter}[\tilde{M}],D_{\rm T}[\hat{N}]\}\,-\,(M\leftrightarrow
N)&=-H_{\rm matter}[\tilde{M}'\hat{N}]\,-\,(M\leftrightarrow N) \nonumber\\
&=H_{\rm matter}[2F_m|E^x|^{1/2}(E^x)'(E^{\varphi})^{-2}(MN'-NM')]\,.
\end{align}
Finally, it is straight forward to check that
\begin{equation}
\{H_{\rm matter}[\tilde{M}],D[\hat{N}]\}\,-\,(M\leftrightarrow N)=\int{\rm d}
x\,(MN'-NM')\frac{2F_{\rm matter}|E^x|^{1/2}(E^x)'}{E^{\varphi}}
\frac{\partial\mathcal{H}_{\rm matter}}{\partial E^{\varphi}}\,.
\end{equation}
Putting everything back in (\ref{HmDmHmDm}),
\begin{align}
\{\tilde{C}_{\rm matter}[M],\tilde{C}_{\rm matter}[N]\}=&\int{\rm d}
x\,(MN'-NM')\bigg[
\frac{|E^x|}{(E^{\varphi})^{2}}\left( \frac{((E^x)')^2}{(E^{\varphi})^2}\bar{\beta}
+4F_{\rm matter}^2\right)\mathcal{D}_{\rm matter}    \nonumber\\
&-\frac{2F_{\rm
    matter}|E^x|^{1/2}(E^x)'}{(E^{\varphi})^2}\left(\mathcal{H}_{\rm
    matter}
-E^{\varphi}\frac{\partial\mathcal{H}_{\rm matter}}{\partial
  E^{\varphi}}\right) \bigg]\,.
\end{align}

\end{appendix}

%\bibliographystyle{../preprint}
%\bibliography{../Bib/QuantGra}

\begin{thebibliography}{10}

\bibitem{SmallLorentzViol}
J.\ Polchinski,
\newblock Comment on [arXiv:1106.1417] ``Small Lorentz violations in quantum
  gravity: do they lead to unacceptably large effects?'', [arXiv:1106.6346]

\bibitem{Regained}
S.~A.\ Hojman, K.\ Kucha\v{r}, and C.\ Teitelboim,
\newblock Geometrodynamics Regained,
\newblock {\em Ann.\ Phys.\ (New York)} 96 (1976) 88--135

\bibitem{ConstraintAlgebra}
M.\ Bojowald, G.\ Hossain, M.\ Kagan, and S.\ Shankaranarayanan,
\newblock Anomaly freedom in perturbative loop quantum gravity,
\newblock {\em Phys.\ Rev.\ D} 78 (2008) 063547, [arXiv:0806.3929]

\bibitem{ScalarHolInv}
T.\ Cailleteau, L.\ Linsefors, and A.\ Barrau,
\newblock Anomaly-free perturbations with inverse-volume and holonomy
  corrections in Loop Quantum Cosmology,
\newblock {\em Class.\ Quantum Grav.} 31 (2014) 125011, [arXiv:1307.5238]

\bibitem{JR}
J.~D.\ Reyes,
\newblock {\em Spherically Symmetric Loop Quantum Gravity: Connections to
  2-Dimensional Models and Applications to Gravitational Collapse},
\newblock PhD thesis, The Pennsylvania State University, 2009

\bibitem{LTBII}
M.\ Bojowald, J.~D.\ Reyes, and R.\ Tibrewala,
\newblock Non-marginal LTB-like models with inverse triad corrections from loop
  quantum gravity,
\newblock {\em Phys.\ Rev.\ D} 80 (2009) 084002, [arXiv:0906.4767]

\bibitem{ModCollapse}
A.\ Kreienbuehl, V.\ Husain, and S.~S.\ Seahra,
\newblock Modified general relativity as a model for quantum gravitational
  collapse,
\newblock {\em Class.\ Quantum Grav.} 29 (2012) 095008, [arXiv:1011.2381]

\bibitem{HigherSpatial}
M.\ Bojowald, G.~M.\ Paily, and J.~D.\ Reyes,
\newblock Discreteness corrections and higher spatial derivatives in effective
  canonical quantum gravity,
\newblock {\em Phys.\ Rev.\ D} 90 (2014) 025025, [arXiv:1402.5130]

\bibitem{ThreeDeform}
A.\ Perez and D.\ Pranzetti,
\newblock On the regularization of the constraints algebra of Quantum Gravity
  in $2+1$ dimensions with non-vanishing cosmological constant,
\newblock {\em Class.\ Quantum Grav.} 27 (2010) 145009, [arXiv:1001.3292]

\bibitem{TwoPlusOneDef}
A.\ Henderson, A.\ Laddha, and C.\ Tomlin,
\newblock Constraint algebra in LQG reloaded : Toy model of a ${\rm U}(1)^{3}$
  Gauge Theory I,
\newblock {\em Phys.\ Rev.\ D} 88 (2013) 044028, [arXiv:1204.0211]

\bibitem{TwoPlusOneDef2}
A.\ Henderson, A.\ Laddha, and C.\ Tomlin,
\newblock Constraint algebra in LQG reloaded : Toy model of an Abelian gauge
  theory -- II Spatial Diffeomorphisms,
\newblock {\em Phys.\ Rev.\ D} 88 (2013) 044029, [arXiv:1210.3960]

\bibitem{AnoFreeWeak}
C.\ Tomlin and M.\ Varadarajan,
\newblock Towards an Anomaly-Free Quantum Dynamics for a Weak Coupling Limit of
  Euclidean Gravity,
\newblock {\em Phys.\ Rev.\ D} 87 (2013) 044039, [arXiv:1210.6869]

\bibitem{SphSymmOp}
S.\ Brahma,
\newblock Spherically symmetric canonical quantum gravity,
\newblock {\em Phys.\ Rev.\ D} 91 (2015) 124003, [arXiv:1411.3661]

\bibitem{SphKl1}
T.\ Thiemann and H.~A.\ Kastrup,
\newblock Canonical Quantization of Spherically Symmetric Gravity in Ashtekar's
  Self-Dual Representation,
\newblock {\em Nucl.\ Phys.\ B} 399 (1993) 211--258, [gr-qc/9310012]

\bibitem{SphKl2}
H.~A.\ Kastrup and T.\ Thiemann,
\newblock Spherically Symmetric Gravity as a Completely Integrable System,
\newblock {\em Nucl.\ Phys.\ B} 425 (1994) 665--686, [gr-qc/9401032]

\bibitem{Kuchar}
K.~V.\ Kucha\v{r},
\newblock Geometrodynamics of Schwarzschild Black Holes,
\newblock {\em Phys.\ Rev.\ D} 50 (1994) 3961--3981

\bibitem{LoopSchwarz}
R.\ Gambini and J.\ Pullin,
\newblock Loop quantization of the Schwarzschild black hole,
\newblock {\em Phys.\ Rev.\ Lett.} 110 (2013) 211301, [arXiv:1302.5265]

\bibitem{LoopSchwarz2}
R.\ Gambini and J.\ Pullin,
\newblock Hawking radiation from a spherical loop quantum gravity black hole,
  [arXiv:1312.3595]

\bibitem{Strobl}
T.\ Strobl,
\newblock Gravity in Two Spacetime Dimensions, [hep-th/0011240]

\bibitem{DiracHamGR}
P.~A.~M.\ Dirac,
\newblock The theory of gravitation in Hamiltonian form,
\newblock {\em Proc.\ Roy.\ Soc.\ A} 246 (1958) 333--343

\bibitem{ADM}
R.\ Arnowitt, S.\ Deser, and C.~W.\ Misner,
\newblock The Dynamics of General Relativity, In L.\ Witten, editor, {\em
  Gravitation: An Introduction to Current Research},
\newblock Wiley, New York, 1962,
\newblock Reprinted in \cite{ADMRe}

\bibitem{CUP}
M.\ Bojowald,
\newblock {\em Canonical Gravity and Applications: Cosmology, Black Holes, and
  Quantum Gravity},
\newblock Cambridge University Press, Cambridge, 2010

\bibitem{NPZRev}
H.\ Nicolai, K.\ Peeters, and M.\ Zamaklar,
\newblock Loop quantum gravity: an outside view,
\newblock {\em Class.\ Quantum Grav.} 22 (2005) R193--R247, [hep-th/0501114]

\bibitem{Energy}
M.\ Bojowald, G.\ Hossain, M.\ Kagan, and C.\ Tomlin,
\newblock Quantum matter in quantum space-time,
\newblock {\em Quantum Matter} 2 (2013) 436--443, [arXiv:1302.5695]

\bibitem{SphSymm}
M.\ Bojowald,
\newblock Spherically Symmetric Quantum Geometry: States and Basic Operators,
\newblock {\em Class.\ Quantum Grav.} 21 (2004) 3733--3753, [gr-qc/0407017]

\bibitem{SymmRed}
M.\ Bojowald and H.~A.\ Kastrup,
\newblock Symmetry Reduction for Quantized Diffeomorphism Invariant Theories of
  Connections,
\newblock {\em Class.\ Quantum Grav.} 17 (2000) 3009--3043, [hep-th/9907042]

\bibitem{LoopSchwarz3}
R.\ Gambini, J.\ Olmedo, and J.\ Pullin,
\newblock Quantum black holes in Loop Quantum Gravity, [arXiv:1310.5996]

\bibitem{GowdyCov}
M.\ Bojowald and S.\ Brahma,
\newblock Covariance in models of loop quantum gravity: Gowdy systems, [in
  preparation]

\bibitem{GowdyAbel}
D.\ Mart\'{\i}n-de Blas, J.\ Olmedo, and T.\ Pawlowski, [in preparation]

\bibitem{SphSymmHam}
B.~K.\ Berger, D.~M.\ Chitre, V.~E.\ Moncrief, and Y.\ Nutku,
\newblock Hamiltonian formulation of spherically symmetric gravitational
  fields,
\newblock {\em Phys.\ Rev.\ D} 5 (1972) 2467--2470

\bibitem{ALMMT}
A.\ Ashtekar, J.\ Lewandowski, D.\ Marolf, J.\ Mour\~ao, and T.\ Thiemann,
\newblock Quantization of Diffeomorphism Invariant Theories of Connections with
  Local Degrees of Freedom,
\newblock {\em J.\ Math.\ Phys.} 36 (1995) 6456--6493, [gr-qc/9504018]

\bibitem{Action}
M.\ Bojowald and G.~M.\ Paily,
\newblock Deformed General Relativity and Effective Actions from Loop Quantum
  Gravity,
\newblock {\em Phys.\ Rev.\ D} 86 (2012) 104018, [arXiv:1112.1899]

\bibitem{SigImpl}
M.\ Bojowald and J.\ Mielczarek,
\newblock Some implications of signature-change in cosmological models of loop
  quantum gravity, [arXiv:1503.09154]

\bibitem{BrownKuchar}
J.~D.\ Brown and K.~V.\ Kucha\v{r},
\newblock Dust as a standard of space and time in canonical quantum gravity,
\newblock {\em Phys.\ Rev.\ D} 51 (1995) 5600--5629

\bibitem{NullDust}
J.\ Bi\v{c}\'ak and K.~V.\ Kucha\v{r},
\newblock Null dust in canonical gravity,
\newblock {\em Phys.\ Rev.\ D} 56 (1997) 4878--4895

\bibitem{RNDeformed}
R.\ Tibrewala,
\newblock Spherically symmetric Einstein-Maxwell theory and loop quantum
  gravity corrections, [arXiv:1207.2585]

\bibitem{LoopReiss}
R.\ Gambini, E.\ Mato~Capurro, and J.\ Pullin,
\newblock Quantum space-time of a charged black hole,
\newblock {\em Phys.\ Rev.\ D} 91 (2015) 084006, [arXiv:1412.6055]

\bibitem{DiscDirac}
V.\ Bonzom and B.\ Dittrich,
\newblock Dirac's discrete hypersurface deformation algebras,
\newblock {\em Class.\ Quantum Grav.} 30 (2013) 205013, [arXiv:1304.5983]

\bibitem{Loss}
M.\ Bojowald,
\newblock Information loss, made worse by quantum gravity,
\newblock {\em Front.\ Phys.} 3 (2015) 33, [arXiv:1409.3157]

\bibitem{QFTCosmo}
A.\ Ashtekar, W.\ Kaminski, and J.\ Lewandowski,
\newblock Quantum field theory on a cosmological, quantum space-time,
\newblock {\em Phys.\ Rev.\ D} 79 (2009) 064030, [arXiv:0901.0933]

\bibitem{AAN}
I.\ Agull\'o, A.\ Ashtekar, and W.\ Nelson,
\newblock An Extension of the Quantum Theory of Cosmological Perturbations to
  the Planck Era,
\newblock {\em Phys.\ Rev.\ D} 87 (2013) 043507, [arXiv:1211.1354]

\bibitem{QSDV}
T.\ Thiemann,
\newblock {QSD V}: Quantum Gravity as the Natural Regulator of Matter Quantum
  Field Theories,
\newblock {\em Class.\ Quantum Grav.} 15 (1998) 1281--1314, [gr-qc/9705019]

\bibitem{ConsistDisc}
R.\ Gambini and J.\ Pullin,
\newblock Canonical quantization of general relativity in discrete space-times,
\newblock {\em Phys.\ Rev.\ Lett.} 90 (2003) 021301, [gr-qc/0206055]

\bibitem{UniformDisc}
M.\ Campiglia, C.\ Di~Bartolo, R.\ Gambini, and J.\ Pullin,
\newblock Uniform discretizations: a new approach for the quantization of
  totally constrained systems,
\newblock {\em Phys.\ Rev.\ D} 74 (2006) 124012, [gr-qc/0610023]

\bibitem{DiffeoOp}
A.\ Laddha and M.\ Varadarajan,
\newblock The Diffeomorphism Constraint Operator in Loop Quantum Gravity,
\newblock {\em Class.\ Quant.\ Grav.} 28 (2011) 195010, [arXiv:1105.0636]

\bibitem{EffAc}
M.\ Bojowald and A.\ Skirzewski,
\newblock Effective Equations of Motion for Quantum Systems,
\newblock {\em Rev.\ Math.\ Phys.} 18 (2006) 713--745, [math-ph/0511043]

\bibitem{EffCons}
M.\ Bojowald, B.\ Sandh\"ofer, A.\ Skirzewski, and A.\ Tsobanjan,
\newblock Effective constraints for quantum systems,
\newblock {\em Rev.\ Math.\ Phys.} 21 (2009) 111--154, [arXiv:0804.3365]

\bibitem{EffConsRel}
M.\ Bojowald and A.\ Tsobanjan,
\newblock Effective constraints for relativistic quantum systems,
\newblock {\em Phys.\ Rev.\ D} 80 (2009) 125008, [arXiv:0906.1772]

\bibitem{CW}
M.\ Bojowald and S.\ Brahma,
\newblock Canonical derivation of effective potentials, [arXiv:1411.3636]

\bibitem{EffConsQBR}
M.\ Bojowald and S.\ Brahma,
\newblock Effective constraint algebras with structure functions,
  [arXiv:1407.4444]

\bibitem{ADMRe}
R.\ Arnowitt, S.\ Deser, and C.~W.\ Misner,
\newblock The Dynamics of General Relativity,
\newblock {\em Gen.\ Rel.\ Grav.} 40 (2008) 1997--2027

\end{thebibliography}

\end{document}